\documentclass[11pt]{article}

\usepackage{osajnl2}
\usepackage{amsmath,bm,bbold}
\usepackage{natbib}

\usepackage{lineno}
\newcommand{\tens}[1]{\mathbf{#1}}
\begin{document}

\title{Optimal polarization modulation and calibration schemes}

\author{R.\ Casini,$^1$ D.\ M.\ Harrington$^2$, and A.\ G.\ de Wijn,$^1$}

\address{${}^1$High Altitude Observatory,
NCAR, P.\ O.\ Box 3000, Boulder, CO 80307-3000, USA\\
${}^2$National Solar Observatory, 22 Ohia Ku Street,
Pukalani, HI 96768, USA}

%\email{${}^*$casini@ucar.edu}

\begin{abstract}
We review the algebraic definition of the efficiency of a 
polarization modulation scheme, which is commonly adopted for solar 
and stellar spectro-polarimetry applications, and generalize 
it to allow distinct states of the modulation cycle to have 
arbitrary throughput and different photon-noise statistics 
for each state. 
Such a generalization becomes necessary to model and optimize 
the polarimetric efficiency of instruments implementing 
spatial polarization modulation schemes, where different 
optical paths are assigned to different polarization analysis 
states, which may be characterized by different throughput values.
The proposed algebraic extension also proves essential for 
introducing a workable concept of the efficiency of a 
polarization calibration scheme, which can then be used to
create a merit function for the optimization of calibration 
sequences, which take into account the specific characteristics of the polarimetric instrument and of its calibration optics.
\end{abstract}

\section{Introduction}
The concept of polarization modulation efficiency is critical both
in the design of highly efficient polarimeters, aimed at optimally
exploiting the photon-flux budget of an instrument, and in devising 
the optimal demodulation scheme, which returns the Stokes parameters 
of the target (and their errors) from a given set of independent, 
modulated intensity signals. The standard reference that introduced 
and analyzed in detail the concept of polarization modulation 
efficiency is \citeauthor{dTC00} (\citeyear{dTC00}, hereafter DTIC; 
see also \citealt{dT03}).

Recently, \cite{dW26} have introduced an analogous concept of 
optimally efficient calibration of a polarimetric instrument, 
which is relevant to improving the performance of calibration 
procedures devised for practical applications. When a polarimetric 
calibration is executed using the science target itself as the 
calibration source (e.g., the Sun), one must guarantee sufficiently 
stable observing conditions of the target, quite similar to the 
conditions required for science observations. This means that 
polarization calibration procedures, though indispensable, take 
up precious time under observing conditions that may be ideal 
for the acquisition of science measurements. The ability to 
design highly performing and minimally redundant calibration 
schemes, which acquire all necessary independent measurements 
in the most optimal way such as to minimize errors on the 
inference of the instrument's modulation matrix, can greatly help 
in reducing the time spent on calibration operations rather than 
science observations. In some cases, this may be vital for preserving 
the calibration system itself, when subject to very intense power fluxes. 
In fact, a preliminary application of the idea of an optimally efficient 
and minimal calibration scheme has been adopted at the NSF Daniel K.\ Inouye Solar Telescope (DKIST), 
motivated by the necessity of keeping the length of exposure of the 
calibration optics to the full solar flux to just a few minutes 
\citep{HS18}.

In this paper we generalize the algebraic definition of polarimetric efficiency 
by relaxing the standard assumption of approximately 
constant signal-to-noise ratio (SNR) during the modulation or calibration 
sequence.  While this might bring just a small correction to most traditional 
polarimetric modulation schemes, when the target is only weakly polarized 
(a quite common occurrence even in solar investigations, perhaps with the 
exclusion of strongly magnetized regions of the solar atmosphere), the 
assumption of approximately constant SNR is certainly inapplicable to the case 
of polarimetric calibration, since the signals reaching the detector virtually 
vary from just below saturation down to the dark level of the detector. 
Even in the case of polarization modulation, however, the throughput of the 
different modulation states can be significantly different in \emph{spatial} 
modulators, where the different signals are channeled into different optical 
systems in parallel. A recent example of a spatial modulator is the 
one developed for the NASA 2020 H-TIDeS Solar Imaging Metasurface Polarimeter 
(SIMPol), where a metasurface polarization splitter (MPS) 
grating \cite[e.g.,][]{Ru21} is used to produce four simultaneous and 
independent states of polarization analysis on one detector, 
sufficient for the inference of the full Stokes vector of the target with every camera frame (\emph{snapshot} polarimetry).

We first present the generalization of the definition of polarization modulation 
efficiency to grasp the basic algebraic manipulation that makes it possible, 
and then apply the same idea to the definition of polarimetric 
calibration efficiency.

\section{Polarization Modulation Efficiency}
\label{sec:modulation}

We consider a general polarimeter with modulation matrix $\tens{O}$.
This is derived from the Mueller matrix of the optical system comprised
between the calibration optics of the system and the polarization
analyzer.
If $n$ is the number of modulation states, then $\tens{O}$ is a
$n\times4$ matrix, and the modulated intensity $n$-vector $\bm{I}$ during the
observation of a target characterized by the Stokes vector\footnote{For notational convenience we identify the Stokes parameters as $S_i$, $i=1,\ldots,4$, where $S_1$ is the total intensity of the target, $S_{2,3}$ measure the two independent states of linear polarization on the plane normal to the propagation direction, and $S_4$ is the amplitude of the circular polarization around that same direction.}
$\bm{S}\equiv(S_1,S_2,S_3,S_4)^t$ is
\begin{equation} \label{eq:modint}
\bm{I}=\tens{O}\bm{S}\;,
\end{equation}
where with $()^t$ we 
indicates matrix transposition. Generally, the transmission efficiencies $t_i$ of the $n$ polarization
``channels'' will be different, so we can define a $n\times n$ \emph{diagonal} transmission 
matrix $\tens{T}=(t_i \delta_{ij})$ and rewrite
eq.~(\ref{eq:modint}) as
\begin{equation} \label{eq:modint1}
\bm{I}=\tens{T}\tens{\tilde O}\bm{S}\;,
\end{equation}
where $\tens{\tilde O}$ is a normalized modulation matrix, which has
the 1st column all made of 1's. 

When $\tens{O}$ is not square---in particular, when the modulation scheme 
consists of more\footnote{A modulation scheme with $n<4$ is evidently 
under-determined for the evaluation of all
four Stokes parameters.} than 4 states---the inversion of 
eqs.~(\ref{eq:modint}) or (\ref{eq:modint1}) generally 
admits an infinite set of possible demodulation matrices $\tens{D}$,
such that
\begin{equation} \label{eq:invert}
\bm{S}=\tens{D}\bm{I}\;.
\end{equation}
The simplest choice---and an optimal one, with regard to the least-square
approximation of the formal solution (\ref{eq:invert})---is provided by the
Moore-Penrose generalized inverse (or \emph{pseudoinverse}) matrix,
$\tens{A}^{\!+}=(\tens{A}^t\tens{A})^{-1} \tens{A}^t$,
where $\tens{A}$ is a generally complex, non-singular matrix \citep{Pe55}. 
The problem of identifying the optimal choice of $\tens{D}$ for 
polarimetric applications has been treated by DTIC. However, 
here we want to relax a major assumption of that work, as described below.

In order to proceed with estimating the statistical noise on the Stokes
parameter $S_i$ implied by the formal solution (\ref{eq:invert}), we 
must first observe that the modulated signals $I_j$ may vary significantly 
depending on the values of the transmissivity $t_j$. Polarization
modulator designs that exclusively employ nearly ideal linear retarders
and polarizers have a $\tens{T}$ matrix that is essentially
proportional
to the identity matrix $\mathbb{1}$. However, this is not a generally
satistifed property of all polarization modulators. For example,
polarimeter designs that perform a \emph{spatial} modulation of the 
incoming Stokes vector by physically directing different modulation 
states into different optical channels, may have a significant 
transmissivity imbalance across the set of modulation states, 
making $\tens{T}$ to measurably depart from a simple rescaling of 
the unit matrix $\mathbb{1}$. In such a case, 
the hypothesis of approximately constant statistical noise of 
the modulated signals, as adopted by DTIC in their treatment, 
is no longer applicable. 

On the other hand, it is generally true that the ratio 
$\tilde I_j=I_j/t_j$ is approximately constant over the set of
modulation states, and proportional to the light source 
intensity $I_0$, \emph{if} such a source is weakly 
polarized, i.e., when $\sum_{i=2}^4S_i^2\ll S_1^2$. While 
the statistical noise of such rescaled signals $\tilde I_j$ 
has no direct physical meaning, under the assumption that 
\emph{all measurements of the modulated signals 
are photon-noise limited}, it is still 
true that, $\sigma^2(I_j)/t_j\approx k I_j/t_j\approx k' I_0$ over the set of modulation states in all polarimeter 
designs, where $k$ and $k'$ are some proportionality factors independent of the
modulation state.\footnote{For example, the measured modulated signals 
$I_j$ may result from some averaging process, either spatial and/or 
temporal, over a number $N$ of nearly identical, Poisson-distributed 
raw signals $I'_j$. In that case, the variance of the mean is 
approximately reduced by $N$, with respect to that of the raw 
signals, i.e., $k=1/N$. Similarly, $k'$ will also include the throughput of the optical system between the light source and the modulator.
\label{fn:nonPoisson}} 
Therefore, we can write, in general,
\begin{eqnarray} \label{eq:noise}
\sigma^2(S_i)
&=&\sum_j D_{ij}^2\,\sigma^2(I_j)
	=\sum_j D_{ij}^2\,t_j\,\sigma^2(I_j)/t_j \nonumber \\
&\approx&\sigma^2_0 \sum_j D_{ij}^2\,t_j
=\sigma^2_0 \sum_j D_{ij}\,t_j D^t_{ji}
=\sigma^2_0\,(\tens{D}\tens{T}\tens{D}^t)_{ii}\;,
\qquad (i=1,2,3,4)
\end{eqnarray}
where in the second line we defined
\begin{equation} \label{eq:sigma0}
\sigma^2_0=\mathrm{avg}\{\sigma^2(I_j)/t_j\}_{j=1,\ldots,n}
\approx\frac{k}{n}\sum_j \frac{I_j}{t_j}\approx
k' I_0\;.
\end{equation}

With the above definitions, the formalism of DTIC for the derivation 
of the optimal demodulation matrix $\tens{D}$ is still applicable, 
as well as the procedure for the estimation of the polarimetric 
efficiencies of the corresponding modulation scheme. 
In particular, the constrained minimization of the noises
(\ref{eq:noise}) described in \S4 of DTIC,\footnote{Such a derivation 
is presented with more detail in the application of our formalism
to the optimization of a calibration sequence, discussed in the 
last section.} yields
\begin{displaymath}
D_{ij}
=\sum_l \lambda_{il}\,O_{jl}/t_j
=\sum_l \lambda_{il}\,O^t_{lj}/t_j
\equiv(\bm{\lambda}\,\tens{O}^t\tens{T}^{-1})_{ij}\;,
\end{displaymath}
where $\bm{\lambda}$ is the matrix of Lagrange multipliers.
The inversion condition $\tens{DO}=\mathbb{1}$ then gives at once\footnote{We note that the symmetry condition $\bm{\lambda}=\bm{\lambda}^t$ is a direct consequence of the symmetry of $\tens{O}^t\tens{T}^{-1}\tens{O}$, along with the commutativity of the operations of matrix transposition and inversion.}
\begin{equation} \label{eq:lambda}
(\bm{\lambda}\,\tens{O}^t\tens{T}^{-1})\tens{O}=\mathbb{1}
\quad\Rightarrow\quad
\bm{\lambda}=(\tens{O}^t\tens{T}^{-1}\tens{O})^{-1}=\bm{\lambda}^t\;,
\end{equation}
and thus the optimal demodulation matrix is
\begin{equation} \label{eq:Ddef}
\tens{D}=(\tens{O}^t\tens{T}^{-1}\tens{O})^{-1}
\tens{O}^t\tens{T}^{-1}\;.
\end{equation}
Evidently, when $\tens{T}=\mathbb{1}$, the above relation simply
gives the pseudoinverse of $\tens{O}$ as demonstrated by DTIC.\footnote{When $\tens{O}$ is square and non-singular, the transmission matrix $\tens{T}$ factors out of Eq.~(\ref{eq:Ddef}), yielding the same demodulation matrix as if $\tens{T}=\mathbb{1}$. Intuitively, this is a necessary consequence of the fact that $\tens{O}$ is invertible, and therefore there exists only one solution for the demodulation matrix, $\tens{D}=\tens{O}^{-1}$. Formally, it is a direct consequence of the theorem on the inverse of a matrix product when all the factor matrices are non-singular; for a product of three matrices it implies $(\tens{ABC})^{-1}=\tens{C}^{-1}\tens{B}^{-1}\tens{A}^{-1}$, which applied to Eq.~(\ref{eq:Ddef}) gives at once $\tens{D}=\tens{O}^{-1}$.\label{fn:theorem}}

The relation (\ref{eq:noise}) then implies that the squares of the \emph{modulation efficiencies} $\epsilon_i$ are inversely proportional to the diagonal elements of
\begin{eqnarray} \label{eq:DtoL}
\tens{D}\tens{T}\tens{D}^t
&=&
\bigl[(\tens{O}^t\tens{T}^{-1}\tens{O})^{-1}
\tens{O}^t\tens{T}^{-1}\bigr]\tens{T}\bigl[\tens{T}^{-1}
\tens{O} (\tens{O}^t\tens{T}^{-1}\tens{O})^{-1}\bigr] \nonumber \\
%&=&
%(\tens{O}^t\tens{T}^{-1}\tens{O})^{-1}
%\tens{O}^t(\tens{T}^{-1}\tens{T})\tens{T}^{-1}
%\tens{O} (\tens{O}^t\tens{T}^{-1}\tens{O})^{-1} \\
&=&
\bigl[(\tens{O}^t\tens{T}^{-1}\tens{O})^{-1}
(\tens{O}^t\tens{T}^{-1}\tens{O})\bigr] 
(\tens{O}^t\tens{T}^{-1}\tens{O})^{-1}
=\bm{\lambda}\;,
\end{eqnarray}
analogously to the results of DTIC.

Sometimes 
%\textbf{\emph{(frankly, I cannot think right now of any
%compelling case for doing so! that's the reason why I decided to invert
%the order of the original derivation...)}}
it may be preferable to work in terms 
of the normalized modulation matrix $\tens{\tilde O}$ defined by
eq.~(\ref{eq:modint1}). 
In such a case, if we indicate with $\tens{\tilde D}$ the 
corresponding pseudoinverse, the inversion of eq.~(\ref{eq:modint1}) 
to determine the input Stokes vector gives
\begin{equation} \label{eq:invert1}
\bm{S}=\tens{\tilde D}\tens{T}^{-1}\bm{I}\;.
\end{equation}
We thus have, for the four Stokes components,
\begin{equation}
S_i=\sum_{jk}\tilde{D}_{ij}(T^{-1})_{jk}\,I_k
=\sum_{j}\tilde{D}_{ij}\,t_j^{-1} I_j\;,
\end{equation}
and the relation (\ref{eq:noise}) for the Stokes errors 
becomes in this case
\begin{eqnarray} \label{eq:noise1}
\sigma^2(S_i)
&=&\sum_j \tilde D_{ij}^2\,t_j^{-2}\,\sigma^2(I_j)
	=\sum_j \tilde D_{ij}^2\,t_j^{-1}\,\sigma^2(I_j)/t_j \nonumber \\
&\approx&\sigma^2_0 \sum_j \tilde D_{ij}^2\,t_j^{-1}
=\sigma^2_0 \sum_j \tilde D_{ij}\,t_j^{-1}\tilde D^t_{ji}
\equiv\sigma^2_0\,
(\tens{\tilde D}\tens{T}^{-1}\tens{\tilde D}^t)_{ii}\;,
\end{eqnarray}
leading to the constrained minimization solution
\begin{displaymath}
\tilde D_{ij}
=\sum_l \lambda_{il}\,\tilde O_{jl}\,t_j
=\sum_l \lambda_{il}\,\tilde O^t_{lj}\,t_j
\equiv(\bm{\lambda}\,\tens{\tilde O}^t\tens{T})_{ij}\;.
\end{displaymath}
The inversion condition $\tens{\tilde D\tilde O}=\mathbb{1}$,
then gives
\begin{equation} \label{eq:lambda.alt}
(\bm{\lambda}\,\tens{\tilde O}^t\tens{T})\tens{\tilde O}=\mathbb{1}
\quad\Rightarrow\quad
\bm{\lambda}=(\tens{\tilde O}^t\tens{T}\tens{\tilde O})^{-1}
=\bm{\lambda}^t\;,
\end{equation}
and the optimal demodulation matrix becomes, in this case,
\begin{equation}
\tens{\tilde D}=(\tens{\tilde O}^t\tens{T}\tens{\tilde O})^{-1}
\tens{\tilde O}^t\tens{T}\;.
\end{equation}
Finally, it is easily verified that, similarly to Eq.~(\ref{eq:DtoL}),
$\tens{\tilde D}\tens{T}^{-1}\tens{\tilde D}^t=\bm{\lambda}$,
and therefore the inverse efficiencies 
$\epsilon_i^{-1}$ are again proportional to the square roots of the diagonal elements of
the $\bm{\lambda}$ matrix (\ref{eq:lambda.alt}), in virtue of Eq.~(\ref{eq:noise1}).

\begin{figure}[t!]
    \centering
    \includegraphics[width=.72\linewidth]{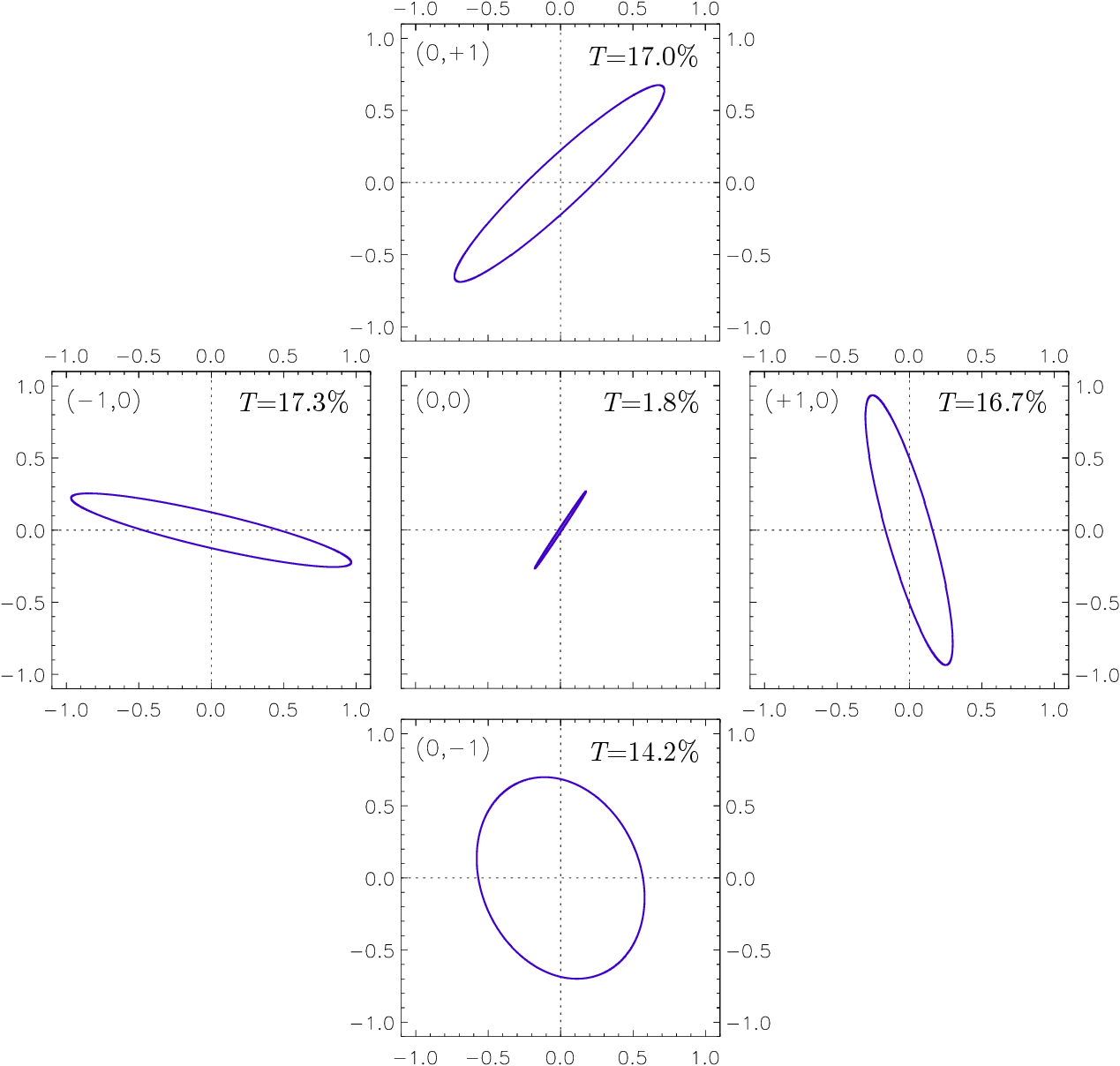}
    \caption{
    %\emph{Top:} Throughput efficiencies of the four diffracted 1st-order channels and the non-diffracted 0th-order channel of the SIMPol MPS, derived from laboratory calibrations. The MPS design optimizes the throughput of the four 1st-order channels. The total throughput of the five channels shown here is ${\sim}\,0.67$. The 0th order dominates the efficiency losses of the MPS grating, along with all the other diffracted channels of order ${>}\, 1$.  Among the four 1st-order channels, the one with the highest throughput one (blue) is 21\% more efficient than the dimmest one (red). \emph{Bottom:} 
    Polarization analysis ellipses of the first five orders of the SIMPol MPS, as derived from laboratory calibrations, along with their percentage throughput. The MPS design optimized the throughput and polarization efficiency of the four 1st-order channels. The total throughput of the five channels shown here is ${\sim}\,67$\%, and the diattenuation values are all above ${\sim}\,97$\%. The (0,0) order dominates the efficiency losses of the MPS grating, along with all the other diffracted channels of order ${>}\, 1$ (not shown).
    %The channels follow the same order as in the bar plot, moving CW from the top-right quadrant.
    The orders shown in the figure follow the same order as in the modulation matrix of Eq.~(\ref{eq:modSIMPol}), moving CW from the (0,+1) order at the top, and ending on the (0,0) order at the center. The reference direction for linear polarization corresponds to the horizontal axis of the plots.
    }
    \label{fig:SIMPol}
\end{figure}

\subsection{Normalized modulation efficiencies}

We conclude this exposition with some remarks about the normalization of the modulation efficiencies based on the number of states in the modulation scheme. As discussed by DTIC, a normalization is necessary in order to be able to compare temporal modulation schemes with different numbers of states. Such a comparison can be made using one of two assumptions: 1) the dwell time per modulation state is the same among different modulation schemes (i.e., same \emph{modulation frequency}; m.f.), or 2) the temporal interval of the modulation cycle is the same among different schemes (i.e., same \emph{modulation period}; m.p.). 

The first assumption corresponds to the case considered by DTIC.
Looking back at Eqs.~(\ref{eq:noise}) and (\ref{eq:sigma0}), when two different modulation schemes $a$ and $b$ are run with the same modulation frequency, $\sigma_0$ is also the same for the two schemes, and thus
\begin{equation} \label{eq:case1}
\frac{\sigma_a^2(S_i)}{\sigma_b^2(S_i)}=
\frac{(\tens{D}_a\tens{T}_a\tens{D}_a^t)_{ii}}{(\tens{D}_b\tens{T}_b\tens{D}_b^t)_{ii}}=
\frac{(\bm{\lambda}_a)_{ii}}{(\bm{\lambda}_b)_{ii}}\;,
\end{equation}
where for the second equivalence we recalled Eq.~(\ref{eq:DtoL}). Because the total number of photons accumulated over a modulation cycle is different for the two modulation schemes, a comparison of the modulation efficiencies in this case is more easily done on a per-state basis. As discussed by DTIC, each state contributes in the mean a statistical noise $n\,\sigma^2(S_i)$ to the inference of the Stokes parameter $S_i$, and so Eq.~(\ref{eq:case1}) leads to
\begin{equation} \label{eq:case1_eff}
\frac{n_a\,\sigma_a^2(S_i)}{n_b\,\sigma_b^2(S_i)}=
\frac{n_a\,(\bm{\lambda}_a)_{ii}}{n_b\,(\bm{\lambda}_b)_{ii}}\equiv\left.\frac{(\epsilon_i)_a^{-2}}{(\epsilon_i)_b^{-2}}\right|_\mathrm{m.f.}\;,
\end{equation}
where for the last equivalence we used the normalization of the modulation efficiencies of DTIC, Eq.~(8).

In contrast, the equality of the modulation period between two different modulation schemes $a$ and $b$ implies (cf.~Eq.~(\ref{eq:sigma0}))
\begin{displaymath}
(\sigma_0^2)_{a,b}\approx \frac{k''}{n_{a,b}}\,I_0\;,
\end{displaymath}
where the new proportionality factor $k''$ is now independent of the particular modulation cycle. 
Additionally, we can account for the distribution of a given incident photon flux across the different number of states of two modulation cycles having the same period by adopting a scaled modulation matrix $\mathbf{O'}=\mathbf{O}/n$. Recalling Eq.~(\ref{eq:lambda}), this implies a correspondingly modified efficiency matrix $\bm{\lambda'}=n^2\,\bm{\lambda}$. The Stokes errors estimated through Eq.~(\ref{eq:noise}) then satisfy the following relation,
\begin{equation} \label{eq:case2_eff}
\frac{\sigma_a^2(S_i)}{\sigma_b^2(S_i)}
%=\frac{n_b}{n_a}\,
%\frac{(\tens{D}_a\tens{T}_a\tens{D}_a^t)_{ii}}{(\tens{D}_b\tens{T}_b\tens{D}_b^t)_{ii}}
=\frac{(\bm{\lambda'}_a)_{ii}/n_a}{(\bm{\lambda'}_b)_{ii}/n_b}
=\frac{n_a\,(\bm{\lambda}_a)_{ii}}{n_b\,(\bm{\lambda}_b)_{ii}}\equiv\left.\frac{(\epsilon_i)_a^{-2}}{(\epsilon_i)_b^{-2}}\right|_\mathrm{m.p.}\;,
\end{equation}
showing that the same normalization of the modulation efficiencies as proposed by DTIC also allows us to compare the \emph{total} errors on the inferred Stokes vector between two different modulation schemes sharing the same modulation period.

\subsection{Applications: SIMPol MPS; Optimization of rotating polychromatic modulators.} \label{sec:modulation.app}

As an example of the application of the concept of generalized modulation efficiency, we consider the case of the SIMPol. The instrument consists of a standalone telescope with a 0.5$^\circ$ field-of-view (FOV) to enable full-disk investigations of solar magnetism. The entrance pupil is mapped onto a MPS grating specifically designed around the 460.7\,nm wavelength of the Sr\,\textsc{I} photospheric line. The MPS is able to channel about 65\% of the incoming photons into the four 1st-order diffraction channels, which have different polarization-analysis characteristics, and are imaged by a field lens into a $2{\times}2$ grid pattern on the detector, with the correspondingly analyzed FOVs fully separated. These enable full-Stokes polarimetry with every camera frame. The non-diffracted 0th-order channel is also polarized, and is imaged at the center of the $2{\times}2$ grid pattern, on the optical axis of the system. Figure \ref{fig:SIMPol} provides a schematic rendition of this arrangement.

The modulation matrix of SIMPol, including the 0th-order channel, was inferred via polarization calibration, and was determined to be
\begin{equation} \label{eq:modSIMPol}
\newcommand{\hd}{\hphantom{-}}
    \tens{O}=\begin{pmatrix}
0.9867 &\hd0.0577 &\hd0.9096 &\hd0.3229 \\
0.9642 &-0.7668 &-0.4640 &\hd0.3021 \\
0.8242 &-0.0968 &-0.0997 &-0.7945 \\
1.0000 &\hd0.8551 &-0.4240 &\hd0.2401 \\
0.1032 &-0.0065 &\hd0.0152 &-0.0058
    \end{pmatrix}\;,
\end{equation}
after normalization by the largest element of the 1st column. 
Figure~\ref{fig:SIMPol} shows the polarization-analysis ellipses of these principal MPS orders along with their throughput values.
We can calculate the modulation efficiencies for the two polarization analysis schemes that employ either just the four channels with the highest throughput or all five channels of SIMPol represented in the modulation matrix (\ref{eq:modSIMPol}), for the inference of the target's Stokes vector. Recalling Eqs.~(\ref{eq:noise}) and (\ref{eq:lambda}), we can simply define the efficiency vector as $\bm{\epsilon}\equiv(\lambda_{11}^{-1/2},\lambda_{22}^{-1/2},\lambda_{33}^{-1/2},\lambda_{44}^{-1/2})$.
In this case, all signals are acquired simultaneously, and therefore there is no ``operational penalty'' associated with employing different numbers of states for the demodulation. Accordingly, we must choose the same normalization for comparing the modulation efficiencies of the two schemes. Because the number of usable states of SIMPol if 5, we choose to adopt that number for the normalization.

For the scheme that only employs the four 1st-order diffraction channels, we find
\begin{displaymath}
\bm{\epsilon}_{\rm DTIC}
\equiv(0.840, 0.514, 0.492, 0.414)\;, \qquad
%\equiv(1.878, 1.150, 1.100, 0.926)\;, \qquad
\bm{\epsilon}_{\rm gen}
\equiv(0.867,0.519,0.496,0.448)\;,
%\equiv(1.940,1.160,1.108,1.001)\;,
\end{displaymath}
respectively using the usual definition of modulation efficiency from DTIC and the generalized definition of this paper. In this case, since $\tens{O}$ is invertible, the demodulation matrices are identical in the two definitions (see footnote \ref{fn:theorem}), so the generalized efficiency vector is only providing a more accurate estimate of the expected SNR on the demodulated Stokes vector. 
Because the design of polarization modulators typically adopts the modulation efficiency as a figure of merit of the optimization procedure, this suggests that differences in throughput among the various modulation states will impact the result of the optimization, because of how such differences affect the determination of the modulation efficiency. In this case, the adoption of the generalized definition of modulation efficiency presented here only serves the purpose of identifying a more optimal modulator design.

When all five modulation states of SIMPol are employed, we find instead
\begin{displaymath}
\bm{\epsilon}_{\rm DTIC}
\equiv(0.841, 0.514, 0.492, 0.414)\;, \qquad
%\equiv(1.881, 1.150, 1.100, 0.926)\;, \qquad
\bm{\epsilon}_{\rm gen}
\equiv(0.880,0.519,0.496,0.448)\;,
%\equiv(1.967,1.161,1.110,1.001)\;,
\end{displaymath}
and the demodulation matrices in this case are also going to be different between the DTIC definition and the generalized one of this paper. In this case, the adoption of the generalized definition leads to a different and more optimal demodulation scheme for the data acquired with a given polarimeter, thus yielding a higher SNR for the demodulated Stokes vectors. This is also predicted by the increased norm of the efficiency vector, most significantly in the case of the generalized demodulation scheme. In fact, the increased efficiency norm for the 5-state modulation scheme demonstrates that, despite the low throughput of the 0th-order channel, the information it carries can still be exploited to decrease the error on the demodulated Stokes vector. In the particular case of SIMPol, because the 0th order is only weakly polarized, this gain is practically realized only on the inference of Stokes $I$. 

The above conclusions are verified by running numerical experiments that simulate the polarization modulation process in the presence of Poisson noise, and by demodulating an ensemble of such synthetic measurements to infer the original Stokes parameters $S_i$ and the associated statistical errors $\sigma(S_i)$ over the ensemble. For the 5-state modulation scheme with matrix (\ref{eq:modSIMPol}), the generalized form of the demodulation matrix is found to produce smaller errors than the DTIC demodulation.

\section{Polarization Calibration Efficiency}
\label{sec:calibration}

The algebraic generalization of efficiency proposed in the previous section can also be applied to the problem of determining
the optimal calibration sequence for a polarization modulator
described by a $n\times 4$ modulation matrix $\tens{O}$ that we wish
to determine \citep{dW26}. 
Similarly to eq.~(\ref{eq:modint}) we can define a $n\times m$ matrix (instead of a $n$-vector) of modulated signals $\mathbf{I}$ produced over a calibration sequence consisting of $m$ measurements,
\begin{equation} \label{eq:calint}
\tens{I}=\tens{O}\tens{C}\;.
\end{equation}
Here $\tens{C}$ is a $4\times m$ matrix, the column vectors of which are the Stokes vectors produced by the calibration optics over the sequence of $m$ calibration states.
 
The problem of optimization of a calibration sequence, therefore,
corresponds to the minimization of the errors on the modulation 
matrix elements $O_{ij}$, with $i=1,\ldots,n$ and $j=1,2,3,4$, that are inferred by the calibration procedure. If we indicate with
$\tens{E}$ one possible pseudo-inverse of $\tens{C}$, then
\begin{equation} \label{eq:calinvert}
\tens{O}=\tens{I}\tens{E}
\quad \Rightarrow \quad
O_{ij}=\sum_{k=1}^m I_{ik}E_{kj}\;.
\end{equation}
Through propagation of errors, assuming that all modulated signals
across the calibration sequence are uncorrelated, we have
\begin{eqnarray} \label{eq:calnoise}
\sigma^2(O_{ij})
&=&\sum_{k=1}^m E_{kj}^2\,\sigma^2(I_{ik})
  =\sum_{k=1}^m E_{jk}^t\,\sigma^2(I_{ik}) E_{kj}\;.
\end{eqnarray}
Since we approximately know the properties of the calibration optics, 
as well as those of the modulator (whether from its design, or 
from some initial guess of a minimization iteration procedure), 
the $n\times m$ modulated signals
\begin{displaymath}
I_{ik}=\sum_j O_{ij}\,C_{jk}
\end{displaymath}
are also approximately known. Therefore, analogously to the 
treatment in Sect.~\ref{sec:modulation}, in particular, under the assumption that 
all measurements of the modulated signals are photon-noise
limited, we can approximate (see also footnote 
\ref{fn:nonPoisson})
\begin{equation}
\sigma^2(I_{ik})/I_{ik}
\approx\mathrm{avg}\{\sigma^2(I_{ik})/I_{ik}\}_{i=1,\ldots,n}^{k=1,\ldots,m}
\equiv\sigma^2_0/I_0\;,
\end{equation}
where the last equivalence stands as the definition of 
$\sigma^2_0$, and where we indicated with $I_0$ the intensity of the 
calibration source (assumed, as before, approximately unpolarized). 
We thus find, from eq.~(\ref{eq:calnoise}),
\begin{equation} \label{eq:calnoise.1}
\sigma^2(O_{ij})
=\sum_{k=1}^m E_{jk}^t\,I_{ik}\,\frac{\sigma^2(I_{ik})}{I_{ik}}\,
	E_{kj}
\approx\frac{\sigma^2_0}{I_0}\sum_{k=1}^m E_{jk}^t\,I_{ik}\,E_{kj}\;.
\end{equation}
If, for each modulation state $i=1,\ldots,n$, we define the $m\times m$ 
diagonal matrix $\tens{T}(i)=(I_{ik} \delta_{lk}/I_0)$, the 
previous equation, for $j=1,2,3,4$, can be rewritten as
\begin{eqnarray} \label{eq:calnoise.2}
\sigma^2(O_{ij})
&\approx&\frac{\sigma^2_0}{I_0}
 \sum_{l,k=1}^m E_{jk}^t\,(I_{ik} \delta_{lk})\,E_{kj} \nonumber \\
&=&\sigma^2_0 \sum_{l,k=1}^m E^t_{jl}\,T_{lk}(i) E_{kj}
\equiv\sigma^2_0\,(\tens{E}^t\,\tens{T}(i)\,\tens{E})_{jj}\;.
\end{eqnarray}

In order to proceed with the optimization of the calibration sequence, 
we impose the minimization of the
total square-error norm over the set of modulation states for each Stokes parameter $S_j$ ($j=1,2,3,4$), 
\begin{eqnarray} \label{eq:calsigma}
\sigma^2_j\equiv\sum_{i=1}^n \sigma^2(O_{ij})
&\approx&\sigma^2_0 \sum_{i=1}^n (\tens{E}^t\,\tens{T}(i)\,\tens{E})_{jj}
=\sigma^2_0\,(\tens{E}^t\sum_{i=1}^n\tens{T}(i)\,\tens{E})_{jj} 
	\nonumber \\
&\equiv&\sigma^2_0\,(\tens{E}^t \tens{\Lambda} \tens{E})_{jj}\;,
\end{eqnarray}
where evidently $\tens{\Lambda}=\sum_{i=1}^n\tens{T}(i)$ is also a
$m\times m$ diagonal matrix.

Following the derivation in \S4 of DTIC for the constrained minimization 
of the above norm, we define, for each Stokes parameter $S_j$,
\begin{eqnarray*}
f_j
&=& (\tens{E}^t \tens{\Lambda} \tens{E})_{jj} -
2\sum_{i=1}^4 \chi_{ji}\left(\sum_{k=1}^m C_{ik}
	E_{kj}-\delta_{ij}\right) \\
&=&\sum_{l,k=1}^m E^t_{jk}\Lambda_{kl}E_{lj} - 
2\sum_{i=1}^4 \chi_{ji}\sum_{k=1}^m C_{ik}
	E_{kj}+2\chi_{jj}\;,
 \qquad (j=1,2,3,4)
\end{eqnarray*}
and impose the vanishing of the gradient $(\partial f_j/\partial E_{pj})$,
for $p=1,\ldots,m$. This gives explicitly
\begin{eqnarray*}
\frac{\partial f_j}{\partial E_{pj}}
&=&\sum_{l,k=1}^m \left(
	\frac{\partial E_{kj}}{\partial E_{pj}}\,\Lambda_{kl}E_{lj} 
	+E_{kj}\Lambda_{kl}\,\frac{\partial E_{lj}}{\partial E_{pj}}
	\right) -
2\sum_{i=1}^4\sum_{k=1}^m \chi_{ji}C_{ik}\,
\frac{\partial E_{kj}}{\partial E_{pj}} \\
&=&\sum_{l,k=1}^m \left(
	\delta_{pk}\Lambda_{kl}E_{lj} 
	+E_{kj}\Lambda_{kl}\delta_{lp}
	\right) -
2\sum_{i=1}^4\sum_{k=1}^m \chi_{ji}C_{ik}\delta_{kp} \\
&=&\sum_{l=1}^m \Lambda_{pl}E_{lj} 
  +\sum_{k=1}^m E_{kj}\Lambda_{kp} -
	2\sum_{i=1}^4\chi_{ji}C_{ip} \\
&=&2\lambda_p E_{pj} -
	2\sum_{i=1}^4\chi_{ji}C_{ip}\;,
\end{eqnarray*}
where in the last line we took into account the diagonality of
the $\tens{\Lambda}$ matrix, recalling its definition,
\begin{equation}
\lambda_p\equiv\Lambda_{pp}=\sum_{i=1}^n I_{ip}/I_0\;,\qquad
p=1,\ldots,m\;.
\end{equation}
The minimization condition then yields
\begin{eqnarray} \label{eq:calE}
&&\lambda_p E_{pj}=
	\sum_{i=1}^4C^t_{pi}\chi^t_{ij}=(\tens{C}^t\bm{\chi}^t)_{pj} \nonumber \\
\Rightarrow\quad
&& E_{pj}=
	\lambda_p^{-1}(\tens{C}^t\bm{\chi}^t)_{pj}\quad\Rightarrow\quad
\tens{D}=\tens{\Lambda}^{-1}\tens{C}^t\bm{\chi}^t
\end{eqnarray}
Since
\begin{equation} \label{eq:UC}
\mathbb{1}=\tens{C E}
=(\tens{C}\tens{\Lambda}^{-1}\tens{C}^t)\bm{\chi}^t
\end{equation}
we finally obtain
\begin{equation} \label{eq:caleff}
\bm{\chi}^t=(\tens{C}\tens{\Lambda}^{-1}\tens{C}^t)^{-1}=\bm{\chi}\;,
\end{equation}
along with the optimal calibration-inversion matrix
\begin{equation} \label{eq:decal}
\tens{E}=\tens{\Lambda}^{-1}\tens{C}^t
	(\tens{C}\tens{\Lambda}^{-1}\tens{C}^t)^{-1}\;.
\end{equation}

The relation (\ref{eq:calsigma}) then implies that the optimization of a calibration sequence corresponds to the 
maximization of the Stokes \emph{calibration efficiencies}, 
which are inversely proportional to the square roots of the diagonal elements of $\bm{\chi}$. In fact, recalling Eqs.~(\ref{eq:calE}) and (\ref{eq:UC}),
\begin{displaymath}
\tens{E}^t\tens{\Lambda}\tens{E}
    =(\bm{\chi}\tens{C}\tens{\Lambda}^{-1})\,\tens{\Lambda}\,(\tens{\Lambda}^{-1}\tens{C}^t\bm{\chi}^t)
    =\bm{\chi}\,(\tens{C}\tens{\Lambda}^{-1}\tens{C}^t)\bm{\chi}^t 
%    &=&\bm{\chi}\mathbb{1}\\
    =\bm{\chi}\;.
\end{displaymath}

Similarly to the case of the modulation efficiency, there is some arbitrariness as to the normalization that can be adopted for expressing the calibration efficiency. For the numerical applications of the next section illustrated by Figs.~\ref{fig:DKIST} and \ref{fig:EliCal}, we defined
\begin{equation} \label{eq:caleff_norm}
\epsilon_i^2\equiv\frac{1}{m^2}\,\mathrm{Tr}(\tens{\Lambda})\,\chi_{ii}^{-1}\;.
\end{equation}
We note that this is numerically equivalent to normalizing the diagonal matrix $\tens{\Lambda}$ by $\mathrm{Tr}(\tens{\Lambda})/m$ prior to calculating $\bm{\chi}$ via Eq.~(\ref{eq:caleff}), and then adopt the usual definition $\epsilon_i^{-2}\equiv m\chi_{ii}$, similarly to Eqs.~(\ref{eq:case1_eff},\ref{eq:case2_eff}).

\subsection{Application: Polarimetric calibrations of SIMPol and DKIST}
\label{sec:calibration.app}

As a demonstration of the calibration optimization formalism, we consider again the case of SIMPol. In an effort to minimize the time spent in lengthy calibration operations of the instrument, we set out to determine
%it is possible to determine 
%a maximally compressed sequence consisting of only 4 
short sequences consisting of a moderately redundant set of angular positions for the combined system of the calibration polarizer and retarder. 
%Such a sequence is strictly sufficient to infer the principal 4-state modulation matrix of SIMPol (or any other 4-state modulator), when all other quantities associated with the calibration optics are known. Otherwise, more redundant calibration schemes become necessary in order to infer all unknown properties of the system.
%
The position angles of the polarizer and retarder for such a sequence with $n$ calibration states are respectively given by
\begin{eqnarray*}
    \alpha(i)=(i-1)\,\Delta\alpha+\delta_\alpha+\varepsilon_\alpha\;, \\
    \beta(i)=(i-1)\,\Delta\beta+\delta_\beta+\varepsilon_\beta\;,
\end{eqnarray*}
%
%for $i=1,2,3,4$,
for $i=1,\ldots,n$,
where $\Delta\alpha$ and $\Delta\beta$ are uniform stepping intervals for the angular positions of the calibration optics across the sequence, $\delta_\alpha$ and $\delta_\beta$ are constant offsets identifying the starting angle of the sequence, and $\varepsilon_\alpha$ and $\varepsilon_\beta$ are possible clocking biases of the mounted optics. %In the case of SIMPol $(\varepsilon_\alpha,\varepsilon_\beta)\equiv(-1.62^\circ,7.10^\circ)$.
The optimization of such a calibration sequence is thus equivalent to determining the most optimal $(\Delta\alpha,\Delta\beta)$ pair, and possibly $(\delta_\alpha,\delta_\beta)$ as well. 

The formalism presented in the previous section was used to code up a non-linear optimization algorithm of the Stokes calibration efficiencies for a modulator with an arbitrary number of states. We then tested the code on the 4-state modulation matrix of SIMPol, i.e., the first four rows of $\tens{O}$ in Eq. (\ref{eq:modSIMPol}). For the sake of demonstration, we assume that both positional biases $\varepsilon_\alpha$ and $\varepsilon_\beta$ of the calibration optics are zero, the waveplate has exactly $\lambda/4$ retardance at the operational wavelength of the instrument (approximately $\pm10$\,nm around the Sr\,\textsc{I} line at 460.7\,nm), and the calibration light source is fully unpolarized (for laboratory calibrations of SIMPol, we used an integrating sphere fed by a blue laser diode). Different assumptions for these quantities will generally make the efficiency optimization converge to different solutions for the calibration sequence.

Table~\ref{tab:SIMPol} reports on the left an optimally efficient and balanced calibration scheme using 7 states. The stepping angles $\Delta$ and offsets $\delta$ of the two calibration optics, as well as the realized angular positions for that sequence are given. We note how the optimization converges towards a solution where the relative positions between the two calibration optics span a symmetric ``fan'' of angles. The approximately converged solution was further optimized to make such a symmetry condition exact, with negligible impact to the maximum efficiency and balance across the polarization Stokes parameters.

\begin{table}[t!]
    \centering
    \begin{tabular}{|c|c|c|c|c|}
    \hline
    \multicolumn{2}{|c}{\textbf{Pol.}} 
    &\multicolumn{3}{|c|}{\textbf{Ret.}} \\
    \hline
    $\Delta$ &$\delta$ &$\Delta$ &$\delta$ &$\rho$ \\
    \hline
    62.11 & \hphantom{-0}0.0\hphantom{0} & 71.52 & -28.23 & \hphantom{1}90.00 \\
    \hline
    \multicolumn{2}{|c}{\emph{pos.}} 
    &\multicolumn{2}{|c|}{\emph{pos.}} &\emph{diff.} \\
    \hline
    \multicolumn{2}{|c}{0.0} &\multicolumn{2}{|c|}{-28.23} &-28.23 \\
    \multicolumn{2}{|c}{62.11} &\multicolumn{2}{|c|}{43.29} &-18.82 \\
    \multicolumn{2}{|c}{124.22} &\multicolumn{2}{|c|}{114.81} &-9.41 \\
    \multicolumn{2}{|c}{186.33} &\multicolumn{2}{|c|}{186.33} &0.0 \\
    \multicolumn{2}{|c}{248.44} &\multicolumn{2}{|c|}{257.85} &9.41 \\
    \multicolumn{2}{|c}{310.55} &\multicolumn{2}{|c|}{329.37} &18.82 \\
    \multicolumn{2}{|c}{12.66} &\multicolumn{2}{|c|}{40.89} &28.23 \\
\hline
    \end{tabular}
   \hspace{12pt}
    \begin{tabular}{|c|c|c|c|c|}
    \hline
    \multicolumn{2}{|c}{\textbf{Pol.}} 
    &\multicolumn{3}{|c|}{\textbf{Ret.}} \\
    \hline
    $\Delta$ &$\delta$ &$\Delta$ &$\delta$ &$\rho$ \\
    \hline
    60.23 & \hphantom{-0}0.0\hphantom{0} & 69.87 & -28.92 & 100.41 \\
    \hline
    \multicolumn{2}{|c}{\emph{pos.}} 
    &\multicolumn{2}{|c|}{\emph{pos.}} &\emph{diff.} \\
    \hline
    \multicolumn{2}{|c}{0.0} &\multicolumn{2}{|c|}{-28.92} &-28.92 \\
    \multicolumn{2}{|c}{60.23} &\multicolumn{2}{|c|}{40.95} &-19.28 \\
    \multicolumn{2}{|c}{120.46} &\multicolumn{2}{|c|}{110.82} &-9.64 \\
    \multicolumn{2}{|c}{180.69} &\multicolumn{2}{|c|}{180.69} &0.0 \\
    \multicolumn{2}{|c}{240.92} &\multicolumn{2}{|c|}{250.56} &9.64 \\
    \multicolumn{2}{|c}{301.15} &\multicolumn{2}{|c|}{320.43} &19.28 \\
    \multicolumn{2}{|c}{1.38} &\multicolumn{2}{|c|}{30.30} &28.92 \\
\hline
    \end{tabular}
    \caption{Two examples of 7-state calibration sequences for the SIMPol instrument, where both calibration polarizer and retarder are always simultaneously in the beam. All values in the table are in degrees. For this optimization, we required that both optics step uniformly according to the corresponding $\Delta$ angle, and starting position $\delta$, the values of which are the results of the calibration optimization. \emph{Left:} optimized sequence for a $\lambda/4$ retarder ($\rho=90^\circ$). \emph{Right:} same sequence after further optimization of the retardance $\rho$ of the calibration retarder. We note how such optimized sequences tend to converge to an angular-domain sampling where the clocking differences between the two optics span a symmetric ``fan'' of angles, centered at $0^\circ$.}
    \label{tab:SIMPol}
\end{table}

\begin{table}[t!]
    \centering
    \begin{tabular}{|c c|c c|c c|}
    \hline
    \multicolumn{2}{|c}{\emph{DKIST current}} 
    &\multicolumn{2}{|c|}{\emph{DKIST relaxed}}
    &\multicolumn{2}{c|}{\emph{DKIST alternate}} \\
    \hline
    \textbf{Pol.} & \textbf{Ret.} &
    \textbf{Pol.} & \textbf{Ret.} &
    \textbf{Pol.} & \textbf{Ret.} \\
    \hline
      0.0 & --- & 0.0 & --- & 16.0 & --- \\
      60.0 & --- & 76.3 & --- & 67.5 & --- \\
      120.0 & --- & 122.2 & --- & 119.0 & --- \\
      0.0 & 0.0 & 0.0 & 0.0 & 16.0 & 0.0 \\
      0.0 & 60.0 & -33.2 & 47.2 & 67.5 & 77.2 \\
      0.0 & 120.0 & -52.4 & 85.6 & 119.0 & 154.4 \\
      45.0 & 30.0 & 20.0 & 54.6 & 170.5 & 231.6 \\
      45.0 & 90.0 & 38.1 & 136.2 & 222.0 & 308.8 \\
      45.0 & 150.0 & 91.9 & 142.0 & 273.5 & 26.0 \\
      45.0 & 0.0 & 57.2 & 7.9 & 325.0 & 103.2 \\
      \hline
    \end{tabular}
    \caption{Three different calibration sequences optimized for the DKIST facility, assuming an optimally efficient and balanced 4-state modulator; all angles are in degrees and are counted CCW looking at the source. \emph{Left:} One of the currently implemented sequences at the DKIST. \emph{Center:} Constrained optimization starting from the sequence in the left panel, using $\pm20^\circ$ initial search interval around the original values, and forcing the first positions of both calibration polarizer and retarder to be zero, as in the original sequence; the optimization was run for the optically contacted SiO$_2$ retarder currently employed at the DKIST. Figure~\ref{fig:DKIST} shows the calibration efficiency curves for the first two sequences in this table. \emph{Right:} Uniformly stepped sequence ($51.5^\circ$ polarizer step, $77.2^\circ$ retarder step, and starting position for the polarizer at $16.0^\circ$) optimized for the DKIST EliCal retarder. Figure~\ref{fig:EliCal} shows the calibration efficiency curves for this last sequence.}
    \label{tab:DKIST}
\end{table}

A further level of optimization can be achieved by allowing the retardance of the calibration waveplate to also be varied, which may be of interest during the design phase of a calibration system. Starting from the 7-state solution derived above, we let both positions and retardance to be optimized, and a new more efficient and better balanced solution is found, still respecting the symmetric-fan structure, and which is reported to the right of Table~\ref{tab:SIMPol}.
%Considering again the second calibration sequence given above with the choice of positive $\delta_\beta$, and forcing this ime $(\Delta\alpha,\Delta\beta)\equiv(45.0^\circ,135.0^\circ)$ exactly, the optimization identifies an optimal retardance of 102.2$^\circ$, along with a slightly modified offset $\delta_\beta=18.1^\circ$.

\begin{figure}[t!]
    \centering
    \includegraphics[width=0.5\linewidth]{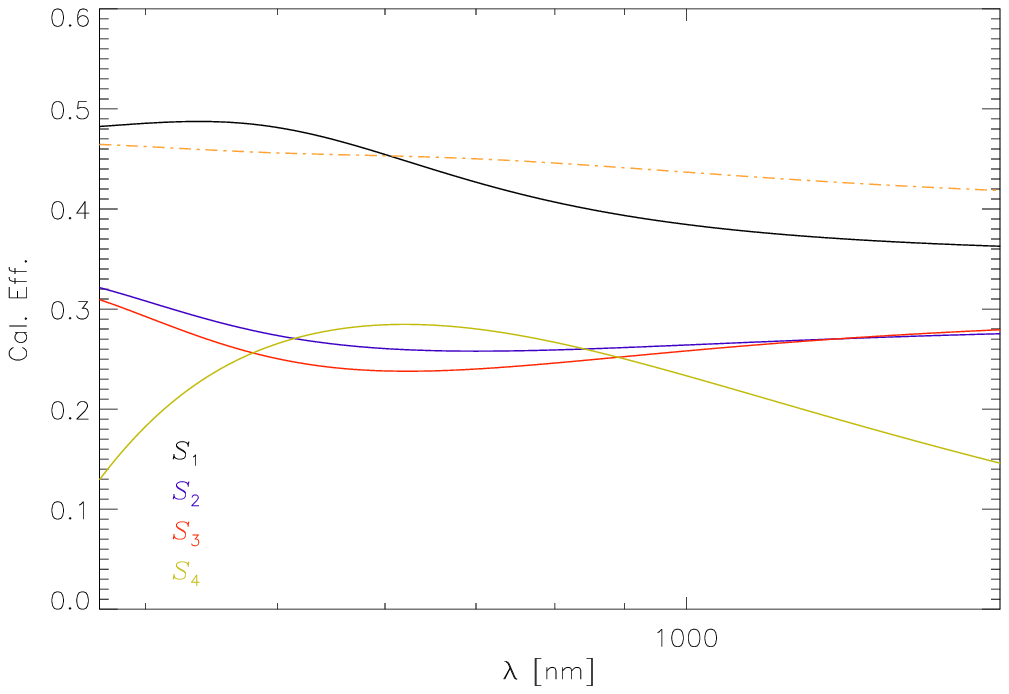}\kern 5pt
    \includegraphics[width=0.5\linewidth]{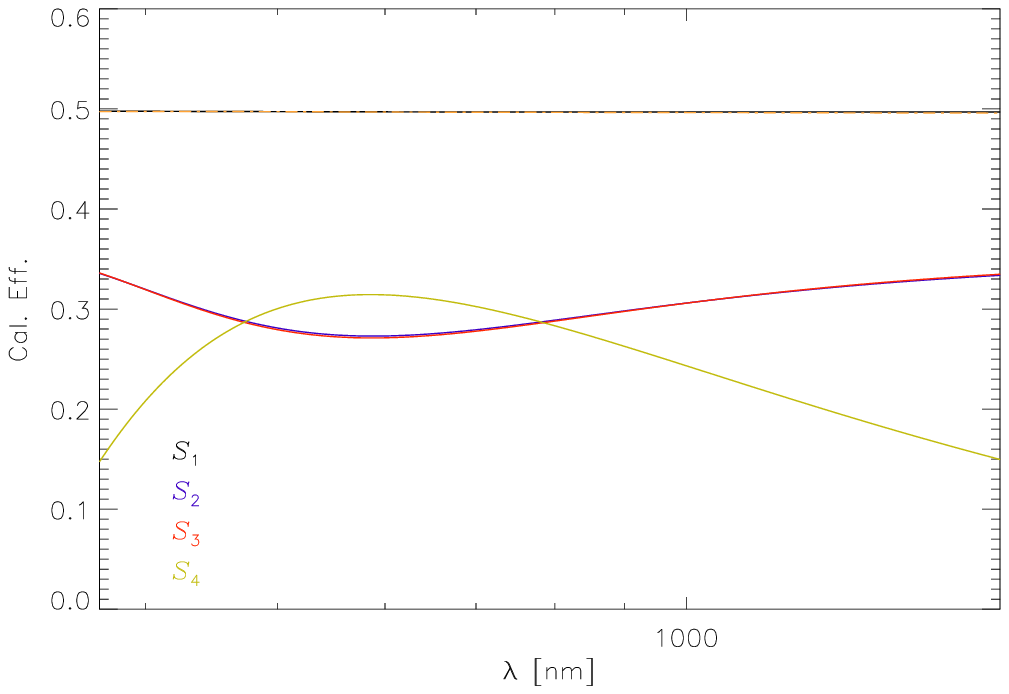}
    \caption{Calibration efficiency plots between 370 and 1700\,nm for the DKIST optically contacted SiO$_2$ calibration retarder, assuming an optimally efficient and balanced 4-state polarization modulator. Colored curves show $S_1$ (black), $S_2$ (blue), $S_3$ (red), and $S_4$ (green). The yellow dot-dashed curve corresponds to the RSS of the $S_{2,3,4}$ efficiencies. \emph{Left:} Using one of the calibration sequences currently implemented at the DKIST (left panel of Table~\ref{tab:DKIST}). \emph{Right:} After optimization of the DKIST sequence, resulting in the calibration sequence in the center panel of Table~\ref{tab:DKIST} (see caption of Table~\ref{tab:DKIST} for details); we note that the curves for the linear polarization parameters $S_2$ and $S_3$ are overlapping, and the RSS of the $S_{2,3,4}$ efficiencies is essentially identical to the efficiency of $S_1$, demonstrating the optimal performance of the calibration sequence for this retarder.}
    \label{fig:DKIST}
\end{figure}

\begin{figure}[t!]
    \centering
\includegraphics[width=0.75\linewidth]{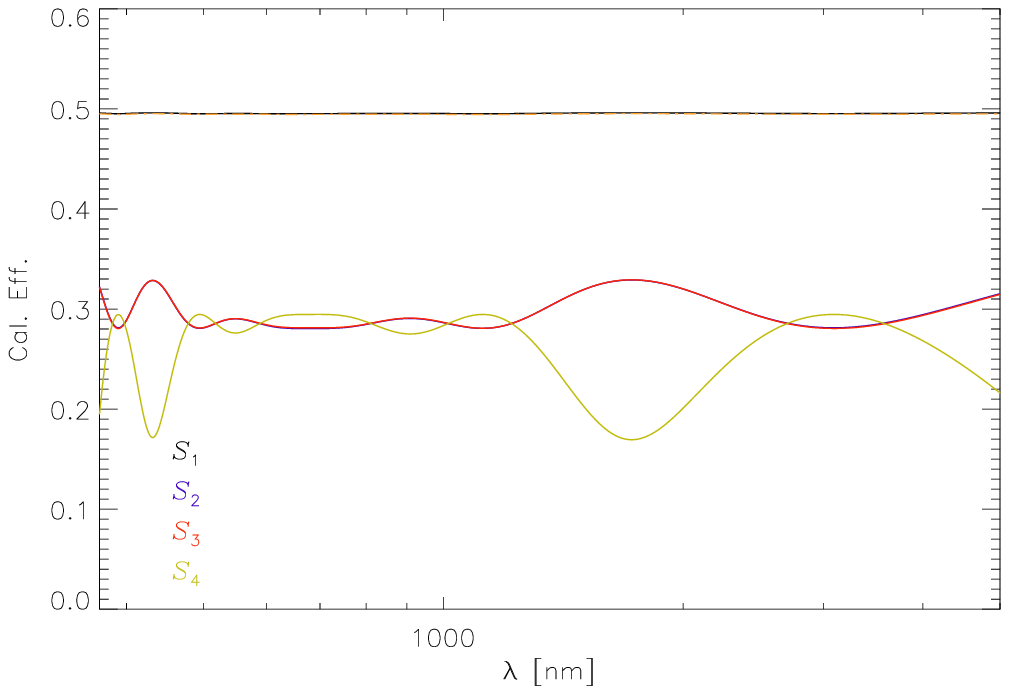}
    \caption{Calibration efficiency plots between 370 and 5000\,nm for the DKIST compound MgF$_2$ elliptical calibration retarder (EliCal), assuming an optimally efficient and balanced 4-state polarization modulator. Colored curves are as in Fig.~\ref{fig:DKIST}. The optimized calibration sequence uses fixed angular steps for both calibration polarizer and retarder, as shown in the right panel of Table~\ref{tab:DKIST}.}
    \label{fig:EliCal}
\end{figure}

The positions of the calibration optics identified by the calibration-efficiency optimization are also dependent on the structure of the calibration sequence itself. For example, if one of the prior sequences
%four-state calibration sequences listed above, using a $\lambda/4$ calibration retarder and a calibration polarizer stepping by $\Delta\alpha=45.0^\circ$, 
is augmented by a series of $n_p$ states where only the polarizer is in the beam (stepping by the same angle $\Delta\alpha$), then the optimization is expected to converge to yet a different set of positions.
%converges to yet a new configuration $(\Delta\alpha;\Delta\beta;\delta_\beta)\equiv(45.0^\circ;134.6^\circ,\{-27.1^\circ,28.1^\circ\})$.

%More generally, one can completely relax the constraint of uniformity of the stepping angles of the calibration optics, and let the optimization procedure directly identify the series $\{\alpha(i),\beta(i)\}_{i,\ldots,m}$ of positional angles that maximizes the efficiencies of the calibration sequence. This is the approach that has been preferred to devise an optimal sequence for the  calibration of the DKIST polarimetric instrumentation. Similarly to the example presented above, degenerate solutions exist that are capable of delivering equivalently optimal and balanced calibration efficiencies, and in the case of an unconstrained search of clocking positions for the two calibration optics, the level of degeneracy is very high, and completely different sequences of positions may be determined starting from an arbitrary random initialization of the optimization problem. This suggests that some initial guesses for the positions should be provided, e.g., motivated by the sensible need to minimize the angular travel of the calibration stages between successive states.

More generally, one can completely relax the constraint of uniformity of the stepping angles of the calibration optics and let the optimization procedure directly identify the set of position angles that maximizes the calibration efficiencies. At DKIST, additional considerations, such as extreme heat loads on the optics and the need to minimize the time-to-SNR ratio with a fairly low calibration duty cycle, drive the optimization towards adopting a minimum number of calibration states. This must achieve a balanced calibration efficiency across $S_{2,3,4}$, greater than some given minimum threshold at all wavelengths, while allowing for the presence of other error sources in the DKIST calibration \citep{Ha20,Ha21,Ha23}. As an example, Fig.~15 of \cite{Ha20} shows a comparison of an oiled 6-crystal super-achromatic retarder (SAR; corresponding to the DKIST original 380-1650\,nm calibration retarder design) with an optically contacted, compound SiO$_2$ zero-order retarder (utilized as the DKIST calibration retarder during commissioning and early operations phases). The latter does have a factor 2.5 worse calibration efficiency spectral balance, with relatively low $S_4$ efficiency at the wavelength extremes. However, this efficiency loss is largely compensated by the fact that spectral fringes are suppressed in this optic (using AR coatings and optical-contact bonding), and heat load impacts are reduced by more than an order of magnitude compared to the SAR-type retarders (reduced spectral clocking oscillations during fabrication, and by using a reduced number of calibration states, always protected by the upstream polarizer). The statistical noise of the calibration process is at least an order of magnitude lower than systematic error limits from other sources so the relatively low calibration efficiency spectral balance is not an issue.

% JATIS8

% DKIST current
%   0.D     0.D
%   60.D    0.D
%   120.D   0.D
%   0.D     0.D
%   0.D     60.D
%   0.D     120.D
%   45.D    30.D
%   45.D    90.D
%   45.D    150.D
%   45.D    0.D
% DKIST optimized (stepped)
%       0.0000000       0.0000000       0.0000000
%       53.427506       0.0000000      -53.427506
%       106.85501       0.0000000      -106.85501
%       0.0000000       77.005051       77.005051
%       53.427506       156.14977       102.72227
%       106.85501       235.29449       128.43948
%       160.28252       314.43922       154.15670
%       213.71002       33.583937       179.87391
%       267.13753       112.72866       205.59113
%       320.56504       191.87338       231.30834
% DKIST optimized (current relaxed)
%        6.5231811       0.0000000      -6.5231811
%       66.048778       0.0000000      -66.048778
%       125.71683       0.0000000      -125.71683
%      -11.664894       20.508995       32.173889
%      -35.112166       72.745317       107.85748
%      -59.672239       124.70196       184.37420
%       69.265700       46.703784      -22.561916
%       41.290194       97.123018       55.832823
%       15.297453       148.83651       133.53906
%       93.897370      -5.1315328      -99.028903

To illustrate the potential of the optimization of calibration efficiencies, we analyzed one of the calibration sequences currently implemented at the DKIST, consisting of 3 configurations involving only the calibration polarizer, followed by 7 configurations with both calibration optics in the beam. Such a sequence is given in left panel of Table~\ref{tab:DKIST}. We started from this sequence, assuming an ideal modulation matrix corresponding to a maximally efficient and balanced modulation scheme \cite[see, e.g., Eq.~(35) of][]{Ca12}, and let it optimize around those initial values to possibly find a better performing sequence. For this exercise, we adopted the model of the compound SiO$_2$ calibration retarder mentioned above. The calibration efficiencies are plotted in Fig.~\ref{fig:DKIST} between 370\,nm and 1700\,nm. The left plot shows the case of the original DKIST sequence, while the right plot shows the efficiency curves for one possible optimal sequence, which is given in the center panel of Table~\ref{tab:DKIST} (see table caption for details). While the performance of the optimized sequence is visibly improved, the qualitative differences are not dramatic, demonstrating that the original sequence already provided good calibration capabilities for this retarder.

%In order to identify improved calibration sequences, we first fixed the polarizer sequence, and let the positions of the retarder to be optimized with an initial search interval of $\pm 20^\circ$ around the nominal values of the sequence. This did not yield a very optimized solution, suggesting that the prescribed positions for the linear polarizer do not optimally probe the parameter space of the system. In fact, a much better solution is found by freezing the retarder positions, and let instead the polarizer positions to be optimized, using the same initial search interval of $\pm 20^\circ$ around the nominal values of the sequence. However, the best optimized sequences were obtained by fully relaxing both polarizer and retarder positions; in particular, this turned out to be essential to achieve a perfect balance between the calibration efficiencies for the linear polarization states (see bottom-right panel of Fig.~\ref{fig:DKIST}). A slightly less optimized sequence, yet significantly improved with respect to the DKIST original one, is obtained by imposing the uniform stepping of both polarizer and retarder, in a similar fashion to the SIMPol example described earlier in this section. The bottom-left panel of Fig.~\ref{fig:DKIST} shows the corresponding calibration efficiency curves.

Figure~\ref{fig:EliCal} shows the result of another calibration sequence optimization, targeting a different type of calibration retarder. This consists of a compound MgF$_2$ elliptical calibration retarder (EliCal), which is planned to be implemented at the DKIST for whole-spectrum calibration of the facility spectro-polarimeters, hence needing to span from 380\,nm to about 4700\,nm in a strictly-simultaneous calibration sequence \cite{Ha23}. The EliCal has obvious limitations determined by the simplicity of its design in contrast with the extremely broad spectral range it must cover. These limitations are clearly demonstrated by the broad dips of the Stokes-$V$ ($S_4$) efficiency around 430\,nm and 1720\,nm. Hence, the efficiency curves shown in Fig.~\ref{fig:EliCal} are representative of the best calibration performance that can be expected with such a simple retarder design. It is important to remark that the calibration efficiencies for the linear polarization states $S_{2,3}$, instead, are not significantly impacted by the retarder design. This is to be expected, as the characterization of the linear-polarization performance of a modulator predominantly relies on the action of the calibration polarizer through the set of angular positions attained along the sequence. The main role of the calibration retarder is to produce circular-polarization states out of the linear-polarization states created by the calibration polarizer, and it only has a secondary effect on the characterization of linear polarization.

This realization allows us to establish a simple principle for predicting the calibration performance of a given retarder design. When we analyze the EliCal as a polarization modulator, and determine its Stokes modulation efficiencies over the spectral range of Fig.~\ref{fig:EliCal}, we see a 1:1 correspondence between the Stokes-$V$ calibration and modulation efficiencies. This immediately suggests that an efficient Stokes-$V$ calibration retarder will be at the same time an efficient Stokes-$V$ modulator. The performance of such a modulator in linear polarization is largely irrelevant for its performance as a full-Stokes calibration retarder. Therefore, one can utilize the modulation-efficiency formalism presented in Sect.~\ref{sec:modulation}, and the optimization methods described, e.g., by \cite{To10}, in order to arrive at a retarder design that is guaranteed in practice to attain nearly optimal and balanced full-Stokes calibration efficiencies. The reverse is also true, and it can be expected that an efficient full-Stokes calibration retarder will also perform efficiently as a Stokes-$V$ modulator. On the other hand, nothing can be said as to the ability of such a retarder to also optimally modulate linear polarization signals.

Similarly to what we demonstrated for the SIMPol case, earlier in this section, it is possible to set up the optimization of the calibration sequence simultaneously with the search of the optical characteristics of the calibration retarder that maximize the resulting efficiencies. This implies expanding the dimensionality of the parameter space sampled by the optimization algorithm, which in the general case of an elliptic retarder requires three additional dimensions.

We conclude this discussion by noting that, for a fully relaxed optimization of a calibration sequence, a virtually infinite set of equally efficient and balanced solutions can be devised. In particular, it is always possible to force one of the positions of the polarizer to be zero, as in the case of the sequence given in the center panel of Table~\ref{tab:DKIST}. This is simply a consequence of the fact that the choice of the reference frame in which the optimal input states of the calibration sequence are specified is completely arbitrary.
%Such one alternate sequence is shown in the rightmost panel of Table~\ref{tab:DKIST}. The corresponding efficiency curves are practically indistinguishable from the ones shown in the bottom-right panel of Fig.~\ref{fig:DKIST}. 

\section{Conclusions}

In this work we generalized the concept of efficiency of a polarization modulation scheme, in order to allow for the possibility that the various modulation states of the polarimeter may produce different enough photon counts that the typical assumption of identical statistical noise of the corresponding modulated signals is no longer applicable. This generalization is achieved through a simple algebraic modification of the least-square optimization problem of the demodulation matrix. We discussed the application of such a generalized formulation to the case of the SIMPol instrument, where the polarization modulation is performed through a metasurface polarization splitter (MPS) grating, which produces distinct and simultaneous polarization analysis channels, corresponding to the various diffraction orders of the grating. Due to instrument design choices, fabrication tolerances, and wavelength dependence of the MPS performance, the usable channels of SIMPol have significantly different throughputs, making the standard definition of modulation efficiency and the derivation of the optimal demodulation matrix \citep{dTC00} no longer  applicable in practice.

The same algebraic generalization can be applied to the problem of determining optimal calibration sequences for the measurement of the modulation matrix of a polarimetric instrument \citep{dW26}. In general, such a matrix may be approximately known from the instrument design, or may have preliminarily been measured in a laboratory, but it may need to be routinely determined or verified under the operational conditions of science observations, with sufficiently small errors to satisfy the polarimetric accuracy requirements of those observations. In this problem, because of the very nature of the polarimetric calibration sequence, which specifically aims at creating large modulation amplitudes of the measured signals across the calibration sequence, the assumption of an approximately constant statistical noise of those signals is completely inadequate. We show how this generalized formalism can effectively used to design optimal calibration sequences given the approximate knowledge of the modulation matrix and the characteristics of the calibration optics, or to re-optimize calibration sequences based on existing optics adopted at major observing facility such as the DKIST.

Finally, we want to comment on the possible interdependence between modulation and calibration efficiency optimizations. While the optimization of an efficient and balanced polarization modulator is completely agnostic about the nature of the calibration optics adopted to measure the modulation matrix in practice, the converse is not generally true. As suggested by Eq.~(\ref{eq:calsigma}), the optimization of a calibration procedure (and of its calibration retarder, if required) formally depends on the type of polarization modulator and its modulation cycle, through the matrix $\mathbf{T}(i)$. For example, in the case of a modulator where the states $i=1,\ldots,n$ are characterized by different throughput values, the creation of the matrix $\mathbf{\Lambda}$ in Eq.~(\ref{eq:calsigma}) will in general be affected by those values. On the other hand, because $\mathbf{\Lambda}$ is practically built as an average over the modulation cycle (see also the discussion at the end of Sect.~\ref{sec:calibration} around Eq.~(\ref{eq:caleff_norm})), one can expect that such a dependence will largely be lifted, and that $\mathbf{\Lambda}$ matrices corresponding to different modulation cycles will approximately differ only by a scaling factor. Under such a reasonable assumption, the optimization of a calibration procedure can safely be performed by applying it to a standard reference modulation cycle, which is what we did in Sect.~\ref{sec:calibration.app} for the analysis of the DKIST calibration sequences.

\bigskip
This material is based upon work supported by the National
Center for Atmospheric Research, which is a major facility
sponsored by the National Science Foundation under Cooperative Agreement No.~1852977.
The authors have greatly benefited from insightful discussions with M.~Collados and J.-C.~del Toro Iniesta on the scope of this work, and on the interpretation of some of the results. R.~C. acknowledges SIMPol collaborators N.~Rubin and L.~Li for helpful discussions on the content of Sects.~\ref{sec:modulation.app} and \ref{sec:calibration.app}, and SIMPol co-I P.~Oakley for the calibration data from which the modulation matrix Eq.~(\ref{eq:modSIMPol}) and Fig.~\ref{fig:SIMPol} were derived.

%\bibliography{geneff}

\begin{thebibliography}{5}
%
\bibitem[\protect\citeauthoryear{Casini, de Wijn, \& Judge}{2012}]{Ca12}
Casini, R., de Wijn, A.~G., \& Judge, P.~G.~2012, ApJ, \textbf{757}, 45 
%
\bibitem[\protect\citeauthoryear{De Wijn et al.}{in prep.}]{dW26}
De Wijn, A.~G., Harrington, D., \& Casini, R., Appl.\ Opt. (in preparation)
%
\bibitem[\protect\citeauthoryear{Del Toro Iniesta \& Collados}{2000}]{dTC00}
Del Toro Iniesta, J.~C., \& Collados, M., Appl.\ Opt., \textbf{39}, 1637 (2000)
%
\bibitem[\protect\citeauthoryear{Del Toro Iniesta}{2003}]{dT03}
Del Toro Iniesta, J.~C., \emph{Introduction to Spectropolarimetry},
Cambridge U.\ Press (2003)
%
\bibitem[\protect\citeauthoryear{Elmore et al.}{2008}]{El08}
Elmore, D.~F., Casini, R., Card, G.~L., Davis, M., Lecinski, A., Lull, R., Nelson, P.~G., \& Tomczyk, S., SPIE, \textbf{7014}, 701416-3 (2008)
%
\bibitem[\protect\citeauthoryear{Harrington \& Sueoka}{2018}]{HS18}
Harrington, D.~M., \& Sueoka, S.~R., JATIS, \textbf{4}, 044006 (2018)
%
\bibitem[\protect\citeauthoryear{Harrington et al.}{2020}]{Ha20}
Harrington, D.~M., Jaeggli, S.~A., Schad, T.~A., White, A.~J., \& Sueoka, S.~R., JATIS, \textbf{6}, 038001 (2020)
%
\bibitem[\protect\citeauthoryear{Harrington}{2023}]{Ha23}
Harrington, D.~M., JATIS, \textbf{9}, 038003 (2023)
%
\bibitem[\protect\citeauthoryear{Harrington et al.}{2021}]{Ha21}
Harrington, D.~M., Schad, T.~A., Sueoka, S.~R., \& White, A.~J., JATIS, \textbf{7}, 038002 (2021)
%
\bibitem[\protect\citeauthoryear{Penrose}{1955}]{Pe55}
Penrose, R., Math.\ Proc.\ Cambridge Phil. Soc., \textbf{51}, 406 (1955)
%
%\bibitem[\protect\citeauthoryear{Rubin et al.}{2022}]{Ru22}
%Rubin, N.~A., Chevalier, P., Juhl, M., Tamagnone, M., 
%Chipman, R., \& Capasso, F., Opt.~Expr., \textbf{30}, 9389 (2022)
%
\bibitem[\protect\citeauthoryear{Rubin et al.}{2021}]{Ru21}
Rubin, N.~A., Shi, Z., \& Capasso, F.~2021, Adv.\ Opt.\ Photon., 13, 836
%
\bibitem[\protect\citeauthoryear{Tomczyk et al.}{2010}]{To10}
Tomczyk, S., Casini, R., de Wijn, A.~G., \& Nelson, P.~G., Appl.\ Opt., \textbf{49}, 3580 (2010)
%
\end{thebibliography}

\end{document}